\documentclass[letterpaper]{article}
\usepackage{graphicx}
\usepackage{aaai}
\usepackage{times}
\usepackage{helvet}
\usepackage{courier}
\usepackage{xcolor}
\usepackage{threeparttable}
\usepackage{dirtytalk}
\usepackage[colorinlistoftodos,prependcaption,textsize=tiny]{todonotes}
\usepackage{amssymb}
\usepackage{subfig}
\usepackage{algorithm}
\usepackage{algpseudocode}

\begin{document}
%
\title{Multi-Gradient Descent for Multi-Objective Recommender Systems}
\author{Nikola Milojkovi\'c,\textsuperscript{\rm 1}\thanks{Work done while at EPFL and Swisscom.} Diego Antognini,\textsuperscript{\rm 2} Giancarlo Bergamin,\textsuperscript{\rm 3} Boi Faltings,\textsuperscript{\rm 2} Claudiu Musat\textsuperscript{\rm 3}\\
\textsuperscript{\rm 1}Oracle Labs, Switzerland\\
\textsuperscript{\rm 2}EPFL, LIA, Switzerland\\
\textsuperscript{\rm 3}Swisscom, Switzerland\\
firstname.lastname@\{oracle.com, epfl.ch, swisscom.com\}
}
\maketitle

\begin{abstract}
Recommender systems need to mirror the complexity of the environment they are applied in. The more we know about what might benefit the user, the more objectives the recommender system has. In addition there may be multiple stakeholders - sellers, buyers, shareholders - in addition to legal and ethical constraints. Simultaneously optimizing for a multitude of objectives, correlated and not correlated, having the same scale or not, has proven difficult so far. 

We introduce a stochastic multi-gradient descent approach to recommender systems (MGDRec) to solve this problem. We show that this exceeds state-of-the-art methods in traditional objective mixtures, like revenue and recall. Not only that, but through gradient normalization we can combine fundamentally different objectives, having diverse scales, into a single coherent framework. We show that uncorrelated objectives, like the proportion of quality products, can be improved alongside accuracy. 
Through the use of stochasticity, we avoid the pitfalls of calculating full gradients and provide a clear setting for its applicability.

\end{abstract}

\section{Introduction}
\paragraph{}
Today, \textbf{recommender systems} are an inevitable part of everyone's daily digital routine. When a person goes online, they are likely going to use one of the services in which recommendation plays an important role. This applies to streaming music, shopping, socializing on social media platforms or viewing a personalized news feed,
among many others. The content is tailored to the user during these activities and is selected by the recommender systems. This automated selection of relevant content makes the entire experience of using digital services more comfortable and engaging \cite{Knijnenburg2012}. Without this, the user would be lost in the enormous and continuously growing quantity of information, products, or choices.

To create the best possible user experience - the most useful recommendations, usually, multiple criteria (often conflicting) have to be taken into account. For example, recommending to users the set of $K$ books, always selected from the set of best-selling books, will not give these users an opportunity to pick some, maybe unpopular, but possibly more appealing book \cite{abdollahpouri2017controlling}. Thus, to reduce the popularity bias in this scenario, the \textbf{additional criteria} would be to create not only relevant but also diverse recommendations.

Regularly the incentives of the service provider are aligned with the satisfaction of the users of their service. However, the provider is also expecting certain benefits from operating the system. For example, one of the objectives of the online retailer might be to increase their profit. Therefore, during the design of such system we also need to take into account the interests of all parties - stakeholders. These may be correlated but they are not necessarily identical and the differences, however subtle, can give rise to tension in converging towards a generally satisfactory solution.

Both of these, additional criteria and stakeholders, increase the level of complexity with which the designers of the recommender system have to deal. In this paper we present a general multi-objective optimization algorithm which resolves these issues. Unlike \cite{burke2018balanced}, we go beyond listing the types of possible objectives and propose a way to jointly optimize them.

\textbf{Multi-objective recommender systems} have long been impractical due to the heavy computation cost involved in the joint optimization. Traditional approaches include evolutionary and genetic algorithms \cite{lin2018multiobjective,lin2019evolutionary,geng2015nnia}. However, these models can only be applied on tiny sets of users and items, which do not scale beyond datasets counting hundreds of samples. This situation does not reflect real use-cases, where we can encounter orders of magnitude more products \cite{Gomez-Uribe:2015:NRS:2869770.2843948} and, similarly, orders of magnitude more users \cite{1167344}.

To alleviate computational issues, other methods optimize each objective sequentially. Generally the accuracy is optimized in a first step such as \cite{di2017adaptive,jugovac2017efficient}, which leads to an initial ranking of the items for a particular user. In a second step, the items are re-ranked using one or more additional objectives. This shortcoming is more visible in problems where the product space is tightly constrained. When few products are available to choose from, if the objectives are not correlated, the rankings for each objective will be materially different. Too few items that are ranked highly for the first objective will also be good solutions for the remaining ones.

To obtain the best of both worlds, existing methods employ a weighted average: either in weighting different losses or in weighting different rankings obtained for each criteria. \cite{ribeiro2015multiobjective} propose a method that combines the output of different algorithms trained for different criteria, and aggregates their ranked lists to provide the final recommendation. 

Finally, \textbf{multi-gradient descent} approaches have been proposed to optimize all objectives simultaneously \cite{Lin:2019:PAM:3298689.3346998}, and provides a better weighting aggregation. In  \cite{Lin:2019:PAM:3298689.3346998} multi-gradient descent is shown to work on correlated objectives. There are however doubts with regard to the mixture of varied objectives and also the applicability on non differentiable functions.

In this work, we address both of these limitations of previous work on multi-objective recommenders based on multi-gradient descent (MGDRec). We first focus on a traditional setting where the optimization is done for two correlated objectives. On two separate datasets, with data on movies and books, we show that MGDRec can create solutions that approach the theoretical optimum - the combination of the best result that can be obtained for each individual objective.

We extend this analysis to show that through gradient normalization and specific training procedures we can extend MGDRec to \textbf{non-correlated objectives}. We focused on recommending unpopular items - documentaries in a movie recommender setup - whose proportion in the recommendation set is anticorrelated with recall. 
Furthermore, we extend the method to non differentiable problems by using the stochastic multi-subgradient descent algorithm (SMSGDA) instead of multi-gradient descent algorithm (MGDA).
Unlike previous work - \cite{Lin:2019:PAM:3298689.3346998}, we \textbf{formalize the use of stochastic optimization} and provide a clear setting for its applicability. 

The main contributions of this paper are:
\begin{itemize}
    \item We find a set of solutions to multi-objective recommendation problems combining varied objectives, using multi-gradient descent. We show that this yields results superior to the state of the art.
    \item We introduce the novel idea of gradient normalization to the multi gradient recommendations. This allows us to combine fundamentally different objectives into the same objective function by using subgradients to relax the differentiability conditions for individual objectives. This flexibility allows us to deal with objectives coming from multiple stakeholders.
\end{itemize}


\section{Related Work}
In  recent  years  a  lot  of  effort  in  recommender  systems  research  is  oriented  towards  improving end-user  experience.  This  led  to  the  increasing  interest  in  objectives  other  than accuracy, and consequently to various approaches in the design of multi-objective recommender systems.

\subsection{Approaches based on Evolutionary Algorithms}

Traditional approaches include evolutionary or genetic algorithms such as Non-dominated Neighbor Immune Algorithm based Recommender System (NNIA-RS) \cite{geng2015nnia}, Probabilistic Multi-Objective Evolutionary Algorithm (PMOEA) \cite{CUI201753}, Non-dominated Sorting Genetic Algorithm II (NSGA-II) \cite{Deb:2002:FEM:2221359.2221582}, Decomposition-based Multi-Objective Evolutionary Algorithm (MOEA/Ds) \cite{lin2019evolutionary} or Multi-Objective Evolutionary Algorithm with Extreme Point Guided (MOEA-EPG) \cite{lin2018multiobjective}.

The main disadvantage of this kind of approach is bad scalability. For example, in \cite{lin2018multiobjective}, the proposed method grows quadratically with the number of users and linearly with the number of items, which leads to a high computational complexity. Therefore, these models can only be applied on tiny sets of users and items, which do not scale beyond datasets counting hundreds of samples.

\subsection{Re-Ranking}

To avoid scalability issues, other works proposed a setup where the recommender is optimized for relevance objective, and then the additional objective is being used for re-ranking. Examples of such methods are Multiple Objective Optimization in Recommendation Systems \cite{Rodriguez:2012:MOO:2365952.2365961}, the greedy strategies of \cite{di2017adaptive} using Maximal Marginal Relevance (MMR) \cite{Vargas:2011:RRN:2043932.2043955}, Explicit Query Aspect Diversification (xQuAD) \cite{vargas2013exploiting}, or user-interaction based \cite{Pu2000,Faltings2004a}, and the post-processing method Personalized Ranking Adaptation (PRA) \cite{jugovac2017efficient}. Also, there is substantial work where the solution is proposed only for a specific objective.

However, these methods apply re-ranking either 1) in a post-processing manner \cite{jugovac2017efficient} or 2) on the top-$n$ recommended items during the optimization process \cite{di2017adaptive}. The former solution leads to sub-optimal solutions, because the recommendation has been trained on a single objective without taking into account others. The latter suffers, in addition, from popularity bias \cite{abdollahpouri2017controlling}, as only the most likely items for a user are reordered.

\subsection{Weighted-Sum of Objectives}

The intuitive and seemingly easy solution to scalability and sub-optimal solutions is to transform the multi-objective problem into a single objective problem. This new single objective would be a weighted sum of all objectives. \cite{ribeiro2015multiobjective} aggregate multiple ranked lists, weights from a graph modeled under a constraint satisfaction problem framework\cite{Torrens2002a}. \cite{Lin:2019:PAM:3298689.3346998} use gradient-descent approaches to provide a better weighting aggregation. Nevertheless, gradient-based methods cannot be applied with non differentiable functions. Finally, the main issue with these method is how to pick an optimal set of weights. In real-world problems doing a grid-search to find these weights might be extremely expensive.

\section{Problem Formulation}
\paragraph{}
In this section we formulate the problem of Multi-Objective Optimization, and define its solution with the Pareto Optimal Solution.

\subsection{Multi-Objective Optimization Problem}
\paragraph{}
A multi-objective optimization problem (MOOP) is an optimization problem in which several possibly conflicting objectives are being optimized simultaneously. It can be defined as follows:

\begin{equation}
\label{eq:opt}
    \min _{w \in R^{D}} \mathbf{L}(\mathbf{w})=\min _{w \in R^{D}}\left|\begin{array}{c}{L_{1}(w)} \\ {L_{2}(w)} \\ {\vdots} \\ {L_{n}(w)}\end{array}\right|_{n \times 1}\
\end{equation}
where $n$ is the number of objectives to optimize, $w$ the model parameters, $D$ the total number of parameter, $L_i: R^{D} \rightarrow R, i=1, \dots, n$, $L_{i}$ is a single objective loss function, and $\mathbf{L}$ the multi-objective loss function. In case that a gradient-based optimization algorithm is applied, the gradient of every constituent loss function $\nabla_{w} L_{i}(w)$ has to be a Lipschitz continuous function \cite{murphy2013machine}.

The operator $min$ in Equation~\ref{eq:opt} represents the operation of minimization of all objectives simultaneously. This is not a limiting factor, because, without loss of generality, any maximization problem can be transformed into a minimization problem.

\subsection{Pareto Optimal Solution}
\paragraph{}

Unlike  the  single-objective  optimization  problems,  in  multi-objective  optimization  problems, in general, there will not exist a unique solution that is better with respect to all objectives. This holds as we cannot make any assumption about the relationship of constituent objectives (whether they are correlated, not correlated, linearly dependent, or independent).  Thus, the solution to the MOOP is not a single solution, but a set of solutions; this set of solutions is called Pareto Set. Before formally defining a concept of Pareto Optimal solution, we first define the concept of Pareto Dominance.
\\
\newline
\textbf{Definition 1}: A solution $w^*$ \textbf{dominates} solution $w$ if for all objectives
$L_i(w^*) \leq L_i(w), i = 1,...,n $ and at least for one objective
$L_j(w^*) < L_j(w), j = 1,...,n $.\\
\newline
\textbf{Definition 2}: A solution $w^*$ is \textbf{Pareto Optimal} if it is \textbf{not dominated} by any other solution $w$.\\
\newline
\textbf{Definition 3}: The set of all \textbf{non-dominated} solutions is called \textbf{Pareto Set}.


\section{Types of Objectives}
\label{sec:objectives}
\paragraph{}

One of the fundamental goals of recommender systems is to create relevant recommendations. Naturally, there is no use in recommending irrelevant items. However, in real-world applications, there are other objectives that we need to satisfy besides relevance: "Good businesses pay attention to what their customers have to say. But what customers ask for and what actually works are very different." \cite{Gomez-Uribe:2015:NRS:2869770.2843948}. In this section, we describe different types of objectives that occurs in recommender systems domain.

\subsection{Semantic Relevance}
\paragraph{}

Semantic relevance, relevance, accuracy, correctness, all of these are different names denoting the same goal of recommending a just-right item to the end-user. In other words, the goal is to recommend an item or a set of items which the end-user is most probably going to like. A tremendous amount of time and effort are spent both in research and industry towards creating new recommender systems that are better than the state-of-the-art with respect to accuracy. Definitely, this is the most important objective in every recommender system. However, it is not the only one we should care about, because the end-user satisfaction is not always correlated to the relevance \cite{Being_Accurate_not_enough}.

\subsection{Correlated to Semantic Relevance}
\paragraph{}

This class of objectives are correlated to the semantic relevance objective. For example, it is not possible to have a \textit{Revenue} bigger than zero if you recommend just irrelevant items. Thus, the \textit{Revenue} is the example of an objective that is correlated with the semantic relevance objective.

\subsection{Not Correlated to Semantic Relevance}
\paragraph{}

Finally, this type of objectives is those that are not correlated to the semantic relevance objective. Contrary to the correlated objectives, here we may have a perfectly \textit{fair} recommendations (if \textit{Fairness} is, for example, an additional objective), and yet these recommendations can be completely irrelevant.


\section{Optimization Algorithm}
To solve the multi-objective optimization problem we will use a gradient-based optimization algorithm. Before presenting the algorithm, we introduce the Common Descent Vectors and present the conditions of optimality for gradient-based solutions in MOOP.

\subsection{Common Descent Vector}
\paragraph{}

A common descent vector is a convex combination of gradients of each objective. It can be defined as follows \cite{desideri:hal-00768935}:

\begin{equation}
    \label{eq:cdv}
    \nabla_{w} \mathbf{L(w)}=\sum_{i=1}^{n} \alpha_{i} \nabla_{w} L_{i}(w)
\end{equation}
where $n$ is number of objectives, $w$ the model parameters, $\nabla_{w} \mathbf{L(w)}$ the common descent vector, $\nabla_{w} L_{i}(w)$ the gradient of the objective function $i$, $\alpha_{i}$ weight of the $i^{th}$ gradient. Equation~\ref{eq:cdv} satisfies the following conditions:

\begin{enumerate}
    \item $\alpha_{1}, \ldots, \alpha_{n} \geq 0$
    \item $\sum_{i=1}^{n} \alpha_{i}=1$
\end{enumerate}

\subsection{Optimality Conditions}
\paragraph{}

In a deterministic multi-objective gradient-based optimization, the necessary conditions for a solution to be optimal are the Karush-Kuhn-Tucker (KKT) conditions. Every solution that satisfies these conditions are called \textbf{Pareto Stationary} \cite{desideri:hal-00768935}.\\
\newline
\textbf{Definition 4}: A solution $w$ is said to be \textbf{Pareto Stationary} if there exists $\alpha_1,...,\alpha_n$ such that:
\begin{enumerate}
    \item $\alpha_1,...,\alpha_n \geq 0$
    \item $\sum_{i=1}^{n} \alpha_{i}=1$
    \item $\sum_{i=1}^{n} \alpha_{i} \nabla_{w} L_{i}(w)=0$
\end{enumerate}
However, the Pareto Stationarity is only the necessary condition of optimality, but not sufficient. The explanation for this can be found if we start from the single objective case: in a single-objective optimization zero gradient is only a necessary condition. It extends to multi-objective problem where the convex combination of gradients is zero, which is just a necessary condition of optimality, but not sufficient.
\\
\newline
\textbf{Definition 5}: Every \textbf{Pareto optimal} solution has to be \textbf{Pareto stationary}, but not every \textbf{Pareto stationary} solution is \textbf{Pareto optimal}.

\subsection{Multi-Gradient Descent Algorithm (MGDA)}
\paragraph{}

The Multi-Gradient Descent Algorithm (MGDA) is an extension of the classical Gradient Descent Algorithm to multiple objectives. This algorithm is proved to converge to the Pareto Stationary solution \cite{desideri:hal-00768935}.

In \cite{desideri:hal-00768935}, the author defines the common descent vector as a minimum L2 norm element in the convex hull of the gradients of each objective. Considering this definition, finding the weights in common descent vector can be formulated as the Quadratic Constrained Optimization Problem (QCOP) \cite{desideri:hal-00768935}.

The QCOP is defined as follows:
\begin{equation}
    \min _{\alpha_{1}, \ldots, \alpha_{n}}\left\{\left\|\sum_{i=1}^{n} \alpha_{i} \nabla_{w} L_{i}(w)\right\|^{2} | \sum_{i=1}^{n} \alpha_{i}=1, \alpha_{i} \geq 0\right\}
\end{equation}{}

After solving QCOP, the common descent vector can be calculated and based on that value, and we either have that:
\begin{itemize}
    \item $\nabla_{w} \mathbf{L(w)}=0$, the solution is Pareto Stationary;
    \item $\nabla_{w} \mathbf{L(w)}\neq0$, the solution is not Pareto Stationary and $\nabla_{w} \mathbf{L(w)}$ is the common descent vector for all objectives.
\end{itemize}

Based on the number of objectives, there are two different ways of how QCOP can be solved: with an analytical solution for two objectives or with a constrained optimization problem for more objectives.

\subsection{Two Objectives}

In case of two objectives the QCOP can be defined as:
\begin{equation}
\label{eq:qcoptwoobjectives}
    \min _{\alpha \in[0,1]}\left\|\alpha * \nabla_{w} L_{1}(w)+(1-\alpha) * \nabla_{w} L_{2}(w)\right\|^{2}
\end{equation}

Then, there is an analytical solution to this problem:

\begin{equation}
\label{eq:alphatwoobjectives}
    \alpha=\frac{(\nabla_{w} L_{2}(w)-\nabla_{w} L_{1}(w))^{T} * \nabla_{w} L_{2}(w)}{\left\|\nabla_{w} L_{1}(w)-\nabla_{w} L_{2}(w)\right\|^{2}}
\end{equation}
where $\alpha$ is clipped to $[0, 1]$.

\subsection{Multiple Objectives}

In case of more than two objectives, we cannot compute an exact solution and have to frame the method under a constrained optimization framework. The efficient solution that scales nicely to the high-dimensional problems is proposed in \cite{NIPS2018_7334}. The proposed solution is based on Frank-Wolfe constrained optimization algorithm~\cite{frank1956algorithm}. In the experiments presented in this paper, the Frank-Wolfe solver is efficient and has excellent convergence properties. Hence, the impact on performance (training time) is insignificant.

\subsection{Stochastic Multi-Subgradient Descent Algorithm (SMSGDA)}
\paragraph{}

Unfortunately, Multi-Gradient Descent Algorithm suffers from multiple drawbacks:
\begin{enumerate}
    \item Calculating full gradient at every optimization step is computationally expensive;
    \item As a deterministic optimization algorithm, it can quickly become stuck at a bad Pareto stationary point; the same way as a full gradient descent algorithm can quickly become stuck at the bad local minimum;
    \item The requirement of calculating gradient for the objective function restricts from using non-smooth loss functions as objective functions (e.g. Mean Absolute Error (MAE)).
\end{enumerate}

These drawbacks limits MGDA from usage in many real-world problems. \cite{poirion:hal-01660788} propose an extension of MGDA to address its limitation, called Stochastic Multi-Subgradient Descent Algorithm (SMSGDA).

The stochasticity in SMSGDA: addresses both computational cost and decreases the probability of being stuck at a bad Pareto stationary point. However, given that we do not calculate full gradient anymore, we are not able to satisfy the third KKT condition in Definition 4: $\sum_{i=1}^{n} \alpha_{i} \nabla_{w} L_{i}(w)=0$. Therefore, we cannot choose the same stopping criteria anymore. Nevertheless, we can use some of the criteria that are regularly used in stochastic optimization problems. For example, we could stop the optimization process if:
\begin{itemize}
    \item a number of epochs has been reached;
    \item the loss of the common descent vector $\mathbf{L(w)}$ is plateauing;
    \item the gradient norm is less than $\epsilon$.
\end{itemize}

In \cite{poirion:hal-01660788}, the authors prove that SMSGDA almost surely converges if non-smooth loss functions are used as objective functions. This alleviates the third drawback of MGDA.

\subsection{Gradient Normalization}
\paragraph{}

When designing recommender systems, we face with various objectives (as described in the Section \ref{sec:objectives}). Additionally, these objectives might have values of the different scales. Both MGDA and SMSGDA are sensitive to different value ranges in objective functions; the gradients would have significantly different norms, leading to the case that one objective completely dominates the whole optimization process.

To alleviate the value range in the multiple objective functions, we propose the following gradient normalization method:
\begin{equation}
    \hat{\nabla_{w}L_i(w)} = \frac{\nabla_{w} L_i(w)}{L_i(w_{init})}
\end{equation}
where $\hat{\nabla_{w}L_i(w)}$ is the normalized gradient vector of a single constituent objective, ${\nabla_{w}L_i(w)}$ the non-normalized gradient vector of a single constituent objective, $L_i(w_{init})$ the initial loss for the particular objective, $w$ the model parameters, and $w_{init}$ the initial parameters of the model. We consider $L_i(w_{init})$ to be an empirical maximum loss for the particular objective. Consequently, the proposed normalization should have the same effect as having an objective loss function that is almost always in the domain $L_i(w) \in [0, 1]$.


\section{Solution selection}

Selecting a Pareto Optimal solution from a Pareto Set is not a trivial task as there is not solution strictly dominating other solutions. However, as a remedy to this problem, there exist several strategies for this task. An overview of the strategies to pick the best solution from a Pareto Set are detailed in \cite{doi:10.1021/acs.iecr.6b03453}. In this paper, we employ the Linear Programming Technique for Multidimensional Analysis of Preference (LINMAP); LINMAP selects an optimal solution based on the Euclidean distance from the ideal point, selecting the solution with the shortest distance \cite{Srinivasan1973}.


\section{Experiments}

\subsection{Datasets}

In order to assess the effectiveness of our proposed model, we first carried out experiments on the well-known \textit{Movielens 20M} dataset\footnote{https://grouplens.org/datasets/movielens/20m/}, and on the \textit{Amazon Books} dataset \cite{mcauley2013hidden}. The second dataset contains multiple target variables that could be used in a multi-objective task. However, the first only includes ratings. We enriched the first dataset by mapping the movies to the \textit{Amazon Movies} dataset \cite{mcauley2013hidden}, and then extracting the prices. Finally, the target variables are ratings, prices and genres for the \textit{Movielens} dataset, and ratings and prices for the \textit{Amazon Books}.

For both datasets, we employed the same preprocessing procedure: we first binarized $5$-star rating (i.e. ratings at three and above are labeled as positive and the rest as negative); users and items were then filtered to those with at least $5$ ratings. A summary of the data we used is shown in Table~\ref{table_data_desc}. Finally, we divided the data into training, validation, test sets corresponding to a split of $90\%$, $5\%$, $5\%$ respectively. Additionnally, we masked out $20\%$ of the items for the validation and test sets.

\begin{table}[!h]
\small
\begin{threeparttable}[t]
    \centering
\begin{tabular}{lcccc} 
Datasets & \#Users & \#Items & \#Interactions & Sparsity\\
\hline
Amazon Books & $93\,976$ & $25\,896$ & $964\,363$ & $99.9604$\\
Movielens\tnote{*} & $132\,580$ & $8\,936$ & $6\,316\,389$ & $99.4669$\\
\end{tabular}
\begin{tablenotes}
     \item[*]Combined with prices from Amazon Movies.
   \end{tablenotes}
\end{threeparttable}
\caption{Dataset statistics (number of users; number of items; number of user-item interactions; sparsity).}
\label{table_data_desc}
\end{table} 

\subsection{Objectives}

In our experiments we focused on two completely different combinations of objectives, the first one was a combination of semantic relevance and revenue, while the second one was a combination of semantic relevance and content quality.

To assess the performance of our model for the semantic relevance objective we use the $Recall@k$ metric to measure the ratio of relevant items that are in the top-k recommendations.

\begin{equation}
\label{eq:recallatk}
    Recall@k(u,\omega) := \frac{\sum_{r=1}^k\mathbb{I}[\omega(r) \in I_u]}{min(k, |I_u|)}
\end{equation}

where $\omega(r)$ denotes the item at rank $r$, $I_u$ is the set of held-out items that user $u$ interacted with and $\mathbb{I}[\cdot]$ is the indicator function. The denominator is the minimum of $k$ and the size of the held-out set $I_u$ \cite{liang2018variational}.

The main goal behind the revenue objective is more or less self explanatory: it is a revenue maximization. The metric we use to measure the performance for this objective is $Revenue@k$. That one is a natural extension of the $Recall@k$ metric; the added component here is the price of a particular item. More precisely, we compute the mean revenue of the top-k relevant recommended items.

\begin{equation}
\label{eq:revenueatk}
    Revenue@k(u,\omega, p) := \frac{\sum_{r=1}^k p(r)*\mathbb{I}[\omega(r) \in I_u]}{min(k, |I_u|)}
\end{equation}
where $p(r)$ is the price of the item at rank $r$. 

Finally, the last objective in our experiments is something that we call content quality. For this one, it is almost impossible to find an universally accepted definition. However, in this context, we refer to content quality as to an objective in which the goal is to increase the amount of \say{content that matters} in the top-k recommendations. The main motivation for choosing that objective is due to the phenomenon known under the name of \say{Filter Bubble} \cite{nguyen2014exploring}. The impact of Filter Bubbles on end-users is that it limits their ability to explore and perceive different content. Who, as a consequence, has a skewed image of the environment. Tackling such a problem requires a lot of effort and interdisciplinary research. However, as a small step forward, we decided to explore the possibility of recommending the \say{content that matters}. Once again, selecting that kind of content is beyond our knowledge. Thus, we decided to name the documentary genre as \say{content that matters} (i.e. content quality). Our goal here was to increase the number of documentaries inside the top-k recommendations, which was at the same time also the metric we used for this objective.

\subsection{Experimental setup}

The approach we propose in this paper is model-agnostic. Therefore the designer of the recommender system has full flexibility in choosing what model he wants to use. For our experiments, we decided to use the Multi-VAE model proposed in \cite{liang2018variational}. The architecture of our model is identical to the one in that paper. The loss function we use for semantic relevance objective is also unchanged.

For the revenue objective we weight the reconstruction loss of the VAE with a price vector which contains the individual prices of each item.

\begin{equation}
\label{eq:revenueloss}
    L_{revenue}(w) = price*L_{reconstruction}(w)
\end{equation}
The intuition behind this loss function is that we want to penalize our model based on the item price. In other words, we want to penalize errors on expensive items more severely.

The loss function for the quality content objective we define as follows:

\begin{equation}
\label{eq:documentaryloss}
    L_{content}(w) = \mathbb{I}_{doc}*popularity*L_{reconstruction}(w)
\end{equation}

where $\mathbb{I}_{doc}$ is an indicator vector that signals if an item is a documentary or not and $popularity$ is the vector of the popularity of each item (e.g. a count of how many users have interacted with an item). Here we want to penalize our recommender more for making mistakes on documentaries, especially on the ones that are more popular.

\subsection{Training Procedure}

The training of the model differs slightly between objectives correlated to semantic relevance and objectives not correlated to semantic relevance. 

The training procedure starts by computing the initial loss of the model for all objectives. We consider these losses to be the empirical maximum losses. Following this step, we train our model by computing the gradients of all objectives over the individual batches of the training set, we normalize them by the empirical maximum losses and compute the $\alpha$ parameters as described in Equation \ref{eq:alphatwoobjectives}. We then use these $\alpha$ parameters to compute the common descent vector illustrated in Equation \ref{eq:qcoptwoobjectives}. And finally, we update the parameters of the model. Algorithm \ref{alg:pseudocode} shows the training procedure in pseudo code.

\begin{algorithm}
    \scriptsize
    \caption{SMSGDA with Gradient Normalization}
    \label{alg:pseudocode}
    \begin{algorithmic}[1] 
            \State $initialize()$
            \For{$i \in 1,...,n$}
                \State $empirical\_loss_i = L_i(w)$
            \EndFor

            \For{$epoch \in 1,...,M$}
                \For{$batch \in 1,...,B$}
                    \State $do\_forward\_pass()$
                    \State $evaluate\_model()$
                    \State $update\_pareto\_set()$
                    \For{$i \in 1,...,n$}
                        \State $calculate\_loss \quad L_i(w)$
                        \State $calculate\_gradient \quad \nabla L_i(w)$
                        \State $normalize\_gradient \quad \nabla\hat{L_i}(w) = \frac{\nabla_wL_i(w)}{empirical\_loss_i}$
                    \EndFor
                    
                    \State $\alpha_1,...,\alpha_n = \textrm{QCOPSolver}\left(\nabla_w\hat{L_1}(w),...,\nabla_w\hat{L_n}(w)\right)$
                    \State $\nabla_wL(w)=\sum_{i=1}^n\alpha_i\nabla_w\hat{L_i}(w)$
                    \State $w = w-\eta\nabla_wL(w)$
                \EndFor
            \EndFor
    \end{algorithmic}
\end{algorithm}

Additional steps we do for the combination of semantic relevance with the non-correlated objectives are that we do not start from the randomly initialized model, but as a starting point, we select the model which was optimized for semantic relevance only. Moreover, in order to nudge the model to learn to recommend items from non-correlated distribution, we inject additional information in end-user preferences. In other words, for the content quality objective, we injected a small positive values for selected content in end-user preference vectors (these values sums up to $1$). While this amount is effective for the model to learn the chosen representation, it is not too large to degrade the performance of semantic relevance. Finally, in order to prevent the model from learning immediately to recommend quality content only, the values of $\alpha$ parameters should be constrained.


\section{Results}

\begin{figure}[!h]
    \centering
    \includegraphics[width=0.475\textwidth]{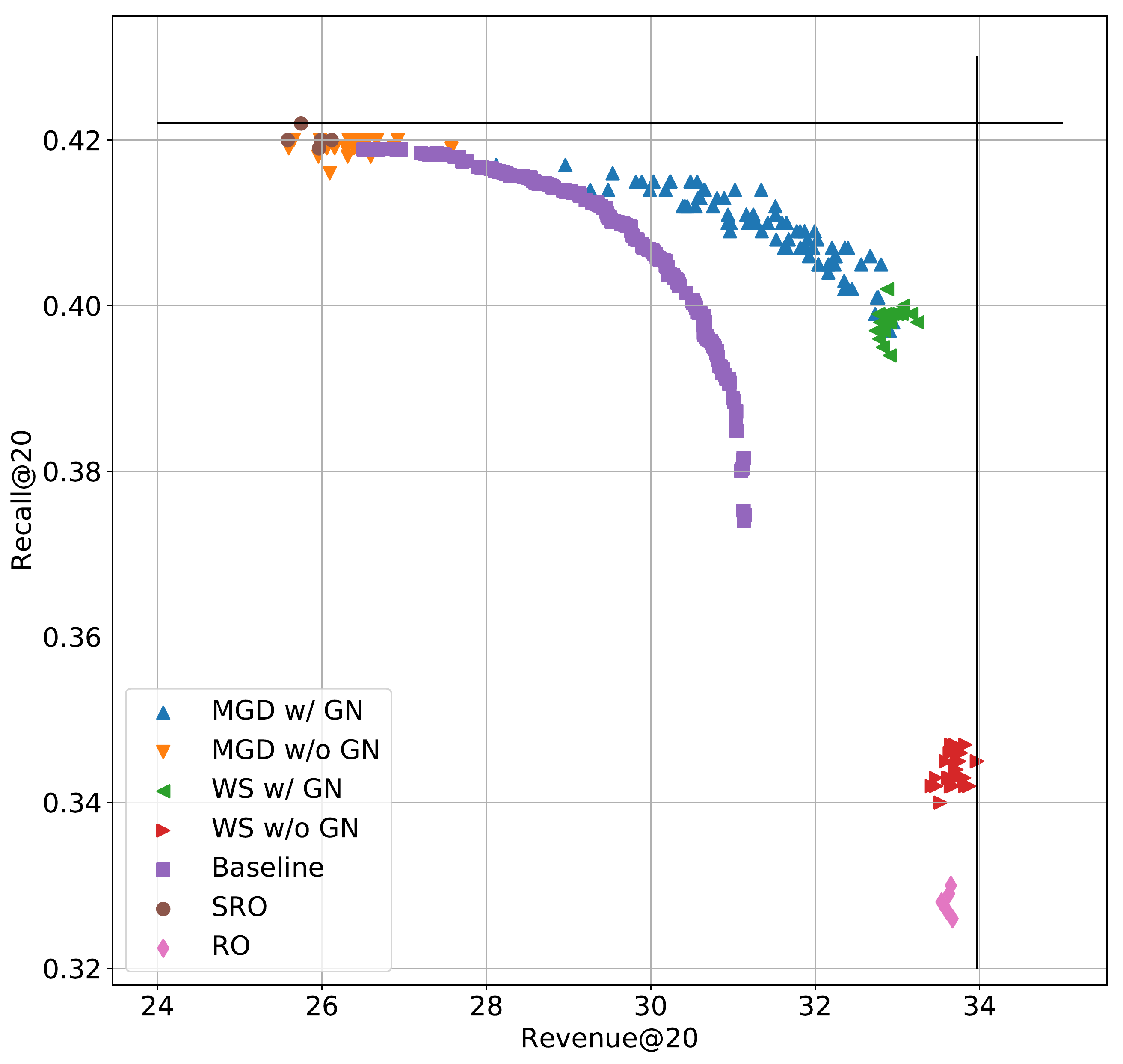}
    \caption{\label{fig:res-movies}Results for two objectives: semantic relevance and revenue, on Movielens dataset combined with Amazon Movies dataset.}
\end{figure}

\begin{figure}[!h]
 \centering
    \includegraphics[width=0.475\textwidth]{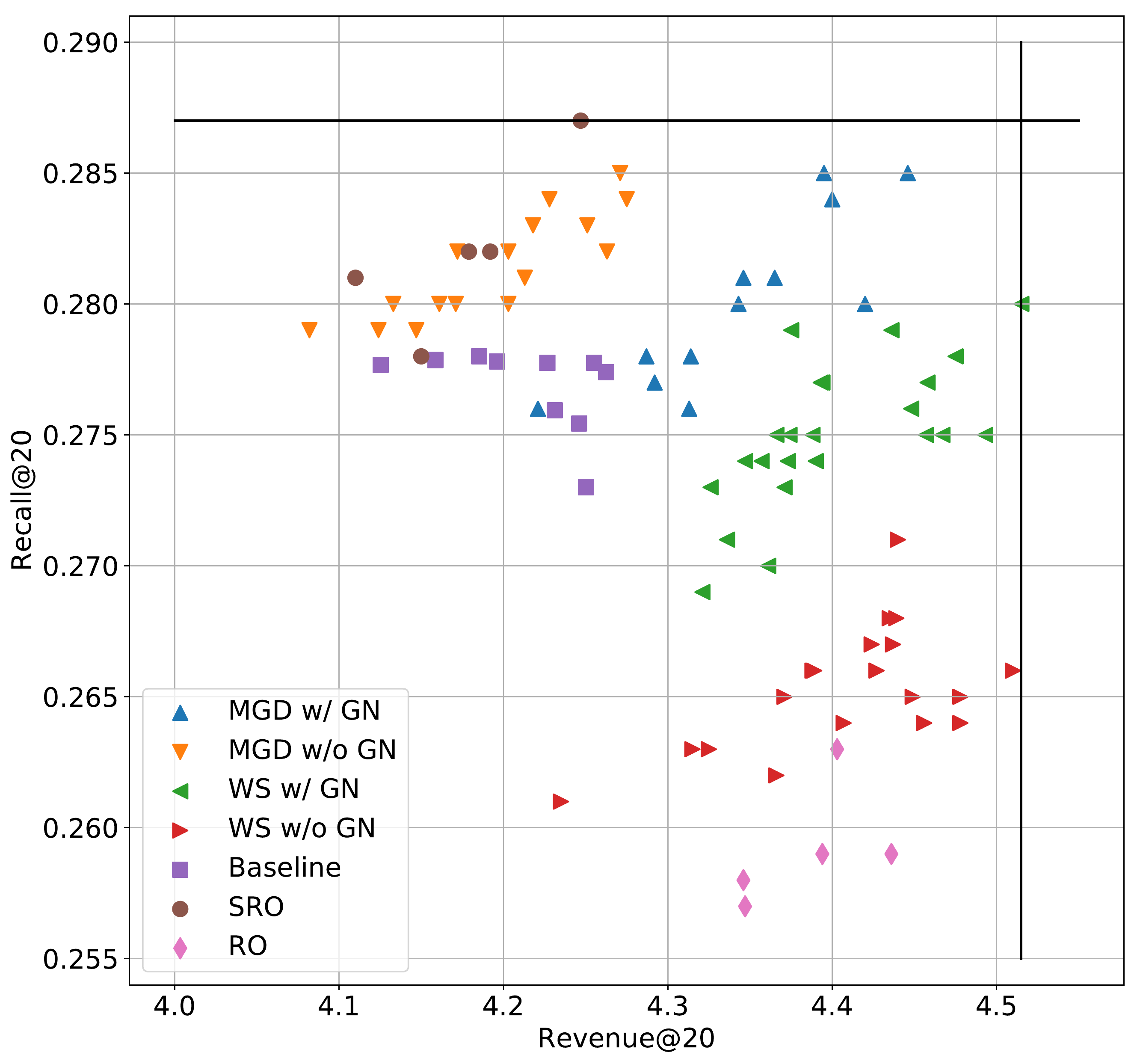}
    \caption{\label{fig:res-books}Results for two objectives: semantic relevance and revenue, on Amazon Books dataset.}
\end{figure}

\newcommand{\comment}[1]{}
\comment{
\textbf{\textcolor{red}{Note:} Results section should be rewritten. There are leftover messages from long time ago.}

As a \textcolor{red}{baseline} we use the closest solution to the presented one \cite{Rodriguez:2012:MOO:2365952.2365961}.
\todo[inline]{GB: I think this baseline part has do be a little more detailed e.g how does the baseline differ from our approach?}
We evaluate the presented algorithm on two different datasets containing reviews of books and movies respectively. In both datasets we merge information about user preference, signaled by reviews and ratings with pricing information. The book dataset is Amazon Books, while the movie dataset is a merger of Movielens Movies with pricing information from Amazon Movies Pricing. 

On both of these, the goal is to combine two objectives: revenue and \textcolor{red}{accuracy}. \todo[inline]{GB: I think we should use consistent naming e.g. relevance \textbf{or} accuracy}
\textcolor{red}{Note}: we will soon finalize the analysis for a separate objective, \textit{diversity}, not directly correlated with either revenue or \textcolor{red}{accuracy}. \todo[inline]{GB: Maybe create a section \say{Future work} and put this there?}We compare the results obtained with variants of the proposed to the state of the art in multi-objective recommendation, as well as with single-objective recommendations.

From the results we have, we can observe that both our methods which use Gradient Normalization found solutions that clearly dominates the considered baseline \cite{Rodriguez:2012:MOO:2365952.2365961}. 

Moreover, our algorithm can also dominate the single-objective baselines, as is the case for the Books experiment, where all the revenue-only solutions are dominated. In all other cases, the tradeoff with respect to the single-objective solutions are appealing, with minimal loss on the common objective and significant gains on the other. For instance, for a big increase in revenue we need to sacrifice little semantic relevance and vice versa -- for small decrease in revenue we have big increase in semantic relevance.

in Figures \ref{fig:res-movies} and \ref{fig:res-books} we show the results obtained with our system and the baselines above. We want the solutions found to be as close as possible to the intersection in the upper right corner - a hypothetical point that reunites the best possible revenue with the best possible accuracy. We notice that using the Euclidean distance in \cite{Rodriguez:2012:MOO:2365952.2365961}, in both cases, the multi-gradient descent with gradient normalization offers the best trade off.

\textbf{New:} In addition, we evaluated our method on yet another set of objective: accuracy(relevance) and increase in number of items from some group(book or movie genre). More concretely, the goal was to increase number of documentaries within top K recommendations. Thus, here we have a combination of two not directly correlated objectives.

In Figure 3 we show that our system was able to significantly increase number of documentaries within recommended items for a price of a small decrease in accuracy. In addition, the baseline was unsuccessful in dealing with two uncorrelated objectives.}


Figure~\ref{fig:res-movies} shows the results for the Movielens dataset enriched with price information from the Amazon Movies dataset. MGD with gradient normalization yields the best results in regard to the solution selection we described above. Figure~\ref{fig:res-books} shows a similar result with MDG with gradient normalization outperforming the other approaches.

To assess how well our algorithm performs we compare it to the other well-known approaches to this problem. As a baseline we use the closest solution to the presented one which was proposed by \cite{Rodriguez:2012:MOO:2365952.2365961} (baseline), which is a solution based on re-ranking. In addition we compare our algorithm also to a simple weighted-sum of objectives in versions with and without gradient normalization (\say{WS w/ GN} and \say{WS w/o GN}). Finally, we compare it also to the models optimized for the single objectives only (\say{SRO} for semantic relevance only and \say{RO} for revenue only).

Both Figures \ref{fig:res-movies} and \ref{fig:res-books} also show that pure MGD without gradient normalization does not perform well, yielding results comparable to the results of the single objective optimization for semantic relevance.
As shown in Figure \ref{fig:res-books} the results of the approaches with gradient normalization outperform the results of the single objective for revenue on the Revenue@k metric. Note that the weighted sum approach also performed significantly better with gradient normalization than without it. This confirms that gradient normalization plays a vital part in increasing the performance of multi objective optimization algorithms. 

\begin{figure}[!h]
    \centering
    \includegraphics[width=0.475\textwidth]{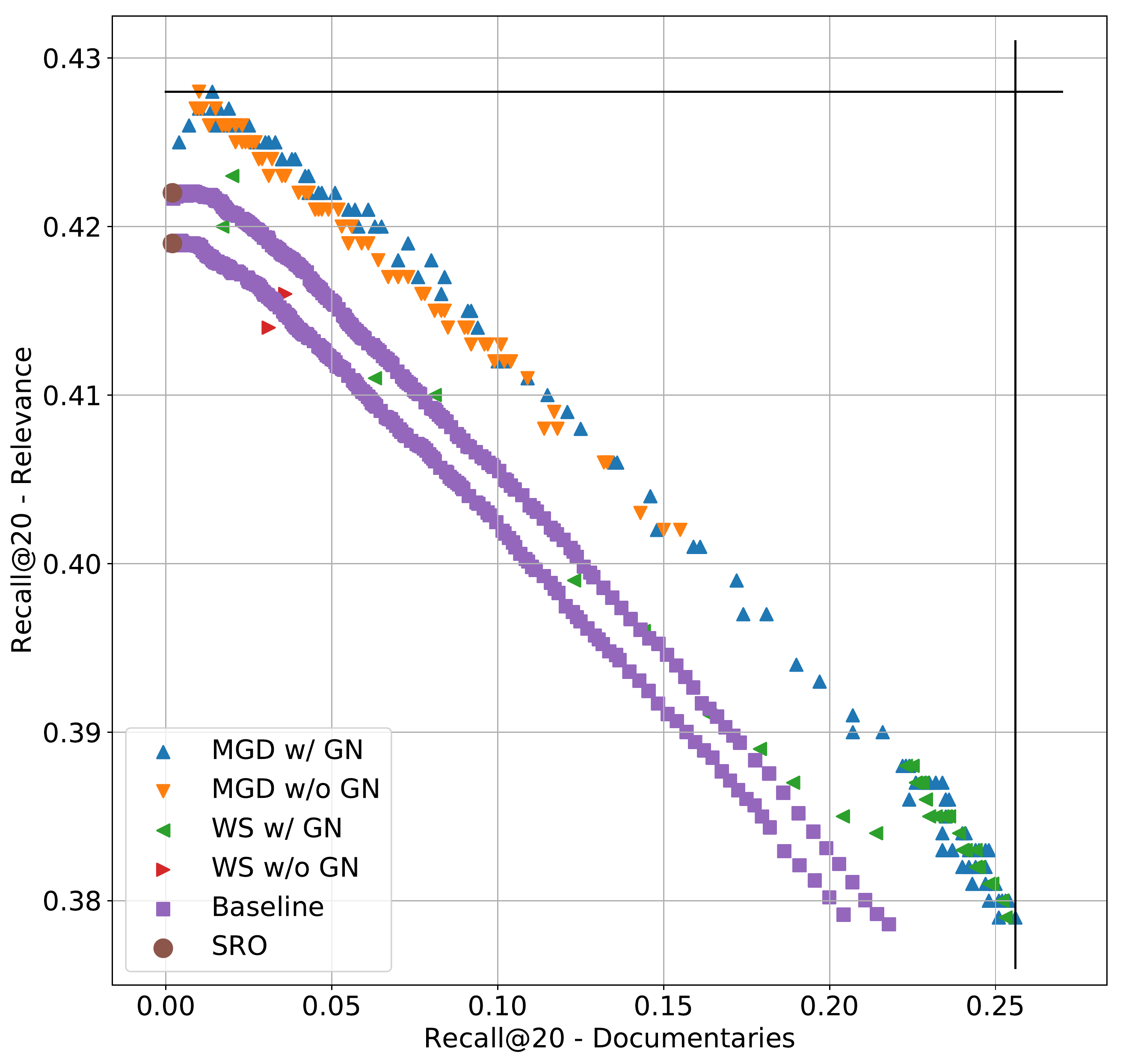}
    \caption{\label{fig:res-documentaries}Results for experiments with two objectives semantic: semantic relevance and content quality, on the Movielens dataset combined with Amazon Movies dataset.}
\end{figure}

Figure \ref{fig:res-documentaries} shows the results of the experiment with the two objectives semantic relevance and content quality. In this experiment we considered movies belonging to the genre \say{documentaries} as quality content. The MGD with gradient normalization dominated the other approaches as well. Interestingly, this approach outperformed also the single objective semantic relevance algorithm on this very objective itself. This may be due to additional regularization provided by the quality content objective.

This result shows that for the cost of a small decrease in semantic relevance we can increase the amount of \say{quality content} in our recommendations drastically.

\section{Conclusion}

The need to optimize for different objectives simultaneously in a recommender system setting is a well recognized problem. While tackling correlated objectives has received more attention previously, it is important for new methods to extend the reach of recommenders to uncorrelated objectives. 
The optimization problem is more complex when the objectives cease to be correlated or are inversely correlated.
In addition, the various objectives may have different scales and may not be differentiable. For these cases, which are actually the norm and stand for real-world concepts like fairness, diversity or revenue, there is a need for novel methods.

In this work we have shown that multi-gradient descent is applicable in this difficult environment. We first tested in a more traditional setup and showed that revenue and recall can be jointly optimized. In two separate experiments, targeting books and movies, we showed that recommenders based on multi-gradient descent (MGDRec) become the new state of the art. 
We then show that completely uncorrelated objectives, like the proportion of a certain type of content - for instance quality or unpopular products - can just as easily be brought into the mix.

We solve the problem of the differences of scale between the objectives using normalization techniques - a novelty that is key for getting the state of the art results. Results show that using the gradient normalization leads to solutions that are the closest to the theoretical optimum - the intersection of the best possible value for each objective taken individually.

\end{document}